\documentclass[twocolumn,times]{aastex631}

\usepackage[utf8]{inputenc}
\usepackage{natbib}
\usepackage{graphicx}
\usepackage{soul}
\usepackage{amsmath}
\usepackage{booktabs}
\usepackage{longtable}
\usepackage{xspace}
\usepackage{hyperref}
\usepackage[T1]{fontenc}
\usepackage{lipsum}
\usepackage{comment}
\usepackage{xcolor}
\usepackage{multirow}
\usepackage{IEEEtrantools}
\usepackage{amsmath}
\usepackage[super]{nth}

\newcommand{\feh}{\ensuremath{\left[{\rm Fe}/{\rm H}\right]}\,}

\newcommand{\teff}{\ensuremath{T_{\rm eff}}\,}

\newcommand{\logg}{\ensuremath{\log g_*}\,}

\newcommand{\prot}{\ensuremath{\,{P_{\rm rot}}}\,}

\newcommand{\mearth}{\ensuremath{\,M_{\oplus}}}

\newcommand{\gaia}{{\it Gaia} }

\newcommand{\mstar}{\ensuremath{M_{*}}}
\newcommand{\rstar}{\ensuremath{R_{*}}}

\newcommand{\vsini}{\ensuremath{v\sin{i_*}}\,}
\newcommand{\exofasttwo}{{\tt EXOFASTv2}\,}
\newcommand{\degree}{\ensuremath{\,^{\circ}}}

\begin{document}
\title{Evidence for Primordial Alignment: Insights from Stellar Obliquity Measurements for Compact Sub-Saturn Systems}

\author[0000-0002-0015-382X]{Brandon T. Radzom}
\affiliation{Department of Astronomy, Indiana University, 727 East 3rd Street, Bloomington, IN 47405-7105, USA}

\author[0000-0002-3610-6953]{Jiayin Dong}
\altaffiliation{Flatiron Research Fellow}
\affiliation{Center for Computational Astrophysics, Flatiron Institute, 162 Fifth Avenue, New York, NY 10010, USA}

\author[0000-0002-7670-670X]{Malena Rice}
\affiliation{Department of Astronomy, Yale University, 52 Hillhouse Ave, New Haven, CT 06511, USA}

\author[0000-0002-0376-6365]{Xian-Yu Wang}
\affiliation{Department of Astronomy, Indiana University, 727 East 3rd Street, Bloomington, IN 47405-7105, USA}

\author[0000-0001-7961-3907]{Samuel W. Yee}
\altaffiliation{51 Pegasi b Fellow}
\affiliation{Center for Astrophysics \textbar \ Harvard \& Smithsonian, 60 Garden Street, Cambridge, MA 02138, USA}
\affiliation{Department of Astrophysical Sciences, Princeton University, 4 Ivy Lane, Princeton, NJ 08544, USA}

\author[0000-0002-0692-7822]{Tyler R Fairnington}
\affiliation{Centre for Astrophysics, University of Southern Queensland, West St,
Darling Heights, Toowoomba, Queensland 4350, Australia}

\author[0000-0003-0412-9314]{Cristobal Petrovich}
\affiliation{Department of Astronomy, Indiana University, 727 East 3rd Street, Bloomington, IN 47405-7105, USA}
\affiliation{Millennium Institute for Astrophysics, Santiago, Chile}

\author[0000-0002-7846-6981]{Songhu Wang}
\affiliation{Department of Astronomy, Indiana University, 727 East 3rd Street, Bloomington, IN 47405-7105, USA}

\begin{abstract}
Despite decades of effort, the mechanisms by which the spin axis of a star and the orbital axes of its planets become misaligned remain elusive. Particularly, it is of great interest whether the large spin-orbit misalignments observed are driven primarily by high-eccentricity migration --- expected to have occurred for short-period, isolated planets --- or reflect a more universal process that operates across systems with a variety of present-day architectures. Compact multi-planet systems offer a unique opportunity to differentiate between these competing hypotheses, as their tightly-packed configurations preclude violent dynamical histories, including high-eccentricity migration, allowing them to trace the primordial disk plane. In this context, we report measurements of the sky-projected stellar obliquity ($\lambda$) via the Rossiter-McLaughlin effect for two sub-Saturns in multiple-transiting systems: TOI-5126\,b ($\lambda=1\pm 48 \degree$) and TOI-5398\,b ($\lambda=-8.1^{+5.3}_{-6.3} \degree$). Both are spin-orbit aligned, joining a fast-growing group of just three other compact sub-Saturn systems, all of which exhibit spin-orbit alignment. In aggregate with archival data, our results strongly suggest that sub-Saturn systems are primordially aligned and become misaligned largely in the post-disk phase, as appears to be the case increasingly for other exoplanet populations.

\end{abstract}

 
\section{Introduction} \label{sec:intro}
Stellar obliquity---i.e., the angle between the spin axis of the host star and the net orbit normal axis of the planets---is a powerful probe of a planetary system's formation history. While the precise mechanisms driving spin-orbit misalignment are not yet clear, it is important to distinguish whether misalignment is largely a product of \underline{high-eccentricity migration}, which should be confined to short-period isolated planets\footnote{Planets that are single or have only distant planetary companion(s)} (e.g., hot Jupiters; see \citealt{Dawson2018} and references therein), or some \underline{universal process} that operates across planetary systems with a variety of architectures, e.g., magnetic warping \citep{Lai2011, Romanova2013}, interactions with stellar companions \citep{Lubow2000, Batygin2012}, stellar flybys \citep{Hao2013}, or stellar gravity waves \citep{Rogers2012, Rogers2013}. 

Compact multi-planet systems may be critical to differentiate between these two scenarios, as their tight orbital configurations are inconsistent with high-eccentricity migration induced by planet-planet scattering \citep{Rasio1996, Chatterjee2008}, Lidov-Kozai cycling \citep{Wu2003, Fabrycky2007, Naoz2016}, or secular interactions \citep{Naoz2011, Wu2011, Petrovich2015} --- all of which tend to disrupt the orbits of nearby planetary neighbors \citep{Mustill2015}. As such, if high-eccentricity migration is the primary driver of misalignment, compact multi-planet systems will retain their primordially aligned configurations. On the other hand, if universal misalignment processes dominate, compact systems may be misaligned at a rate comparable to isolated systems. 

The great majority of spin-orbit angle measurements have been made for close-in Jupiters, particularly hot Jupiters (see recent review by \citealt{Albrecht2022}), which rarely host nearby companions \citep{steffen2012, Huang2016, Hord2021, Wu2023}. Therefore, it is useful to expand the census of stellar obliquities to other types of exoplanets. Sub-Saturns, broadly defined as giant planets with masses ranging between that of Neptune ($\sim$17 M${\oplus}$) and Saturn ($\sim$95 M${\oplus}$), represent a particularly interesting population for such studies. The core accretion model implies that they are `failed gas giants' \citep{Pollack1996}, suggesting they have analogous origins to Jupiters \citep{Dong2018, Lee2019, Hallatt2022}. However, sub-Saturns are less amenable to detection than Jupiters and are relatively understudied as an exoplanet population, so few constraints on their formation histories exist. As of this writing, only about two dozen sub-Saturns have had their spin-orbit angle measured, demonstrating that these planets can take on a range of values from near-perfect alignment to almost completely retrograde. Yet, only \emph{three} of these measurements have been made for sub-Saturns in compact multi-planet systems: Kepler-9\,b \citep{Wang2018}, AU\,Mic\,b \citep{Hirano2020b}, and WASP-148\,b \citep{Wang2022}. While this sample is limited in number, all three systems show evidence of spin-orbit alignment and thus provide tentative support for primordial alignment.

The Transiting Exoplanets Survey Satellite (TESS) mission \citep{Ricker2015}, now well into its Extended Mission phase, has produced a growing sample of sub-Saturns around bright stars. Of particular interest to this work, TESS is responsible for the discovery of at least two sub-Saturns in compact multi-planet systems: TOI-5126\,b \citep{Fairnington2023}, and TOI-5398\,b \citep{Mantovan2022, Mantovan2024a}.
Given the brightness of their host stars and advances in Extreme Precision Radial Velocity (EPRV), these systems are amenable to stellar obliquity measurements.

In this study, we present Rossiter-McLaughlin (RM) effect \citep{Holt1893,Rossiter1924,McLaughlin1924,Queloz2000} measurements for both confirmed compact TESS sub-Saturn systems, TOI-5126 and TOI-5398, using the high-precision NEID spectrograph \citep{Schwab2016} on the WIYN $3.5\, \mathrm{m}$ telescope at Kitt Peak, AZ.
TOI-5126, a relatively bright ($V=10.1 \, \mathrm{mag}$) late-type F dwarf ($\teff=6297\,\mathrm{K}$), hosts a hot sub-Saturn with a 5.5-day period and an outer warm sub-Neptune with a 17.9-day period \citep{Fairnington2023}. 
TOI-5398, a bright ($V=10.1 \, \mathrm{mag}$) early-type G dwarf ($\teff=6039\,\mathrm{K}$), harbors a warm sub-Saturn with a 10.6-day period and an inner hot sub-Neptune with a 4.8-day period \citep{Mantovan2022,Mantovan2024a}. Our reported spin-orbit angles for these two systems represent a significant contribution to the limited sample of sub-Saturns in compact multi-planet systems with such measurements. This work is additionally part of the 11th published outcome of the Stellar Obliquities in Long-period Exoplanet Systems (SOLES) survey, as detailed in a series of publications (\citealt{Rice2021}; \citealt{Wang2022}; \citealt{Rice2022b}; \citealt{Rice2023a}; \citealt{Hixenbaugh2023}; \citealt{Dong2023}; \citealt{Wright2023}; \citealt{Rice2023b}; \citealt{Lubin2023, Hu2024}).

This paper is organized as follows. In Section \ref{sec:obs}, we outline our radial velocity observations. In Section \ref{sec:stellar}, we describe our determination of stellar parameters. In Section \ref{sec:obliq}, we detail our modeling of the RM signals and, subsequently, the stellar obliquities. In Section \ref{sec:discussion}, we quantify the significance of our results and discuss their implications for the origins of spin-orbit misalignment and the formation histories of sub-Saturns.

\section{Observations}\label{sec:obs}

The RM effect of TOI-5126 and TOI-5398 were observed using the WIYN/NEID spectrograph in the High Resolution (HR) mode ($R \sim 110,000$, \citealt{Halverson2016,Schwab2016}) on February 28, 2023 and April 16, 2023, respectively.
NEID is a fiber-fed \citep{Kanodia2018}, actively environmentally stabilized spectrograph \citep{Stefannson2016, Robertson2019} with a wavelength coverage of 380\,nm to 930\,nm.
For TOI-5126\,b, we obtain 20 radial velocity (RV) measurements in HR mode with 1000-second exposures from 02:20--07:48 UT. In total, 14 of these RVs cover the full transit while the remaining six provide out-of-transit baseline; one occurring $0.3\,\mathrm{hr}$ pre-ingress and the other five spanning nearly $1.5\,\mathrm{hr}$ post-egress. These observations occurred under atmospheric conditions with a seeing range of 1.2$\arcsec$--2.8$\arcsec$ (median 1.75$\arcsec$) and an airmass range of $\mathrm{z} = 1.03$--$2.31$. At a wavelength of $5500 \text{\AA}$, the NEID spectrograph achieved a signal-to-noise ratio (SNR) of 56 pixel$^{-1}$. 
For TOI-5398\,b, we obtain 14 RV measurements in HR mode with 1250-second exposures from 02:55--09:04 UT, eight of which were in-transit, sampling nearly the entire transit (excluding ingress). The conditions featured a seeing range of 0.7$\arcsec$--1.5$\arcsec$ (median 1.0$\arcsec$) and an airmass range of $\mathrm{z} = 1.08$--$1.71$. At the same wavelength, the spectrograph's SNR was 60 pixel$^{-1}$.
We obtain an additional 14 NEID RVs covering TOI-5398\,b's orbit interspersed between April 8, 2022 and December 27, 2023, four with 240-second exposure times and nine with 300-second exposure times, and we utilize these measurements in subsequent global fitting for this system.

The NEID spectra were analyzed using version 1.3.0 of the NEID Data Reduction Pipeline (\texttt{NEID-DRP})\footnote{Detailed information is available at: \url{https://neid.ipac.caltech.edu/docs/NEID-DRP/}}, and radial velocities were derived via the cross-correlation function (CCF) method within \texttt{NEID-DRP}. We extracted the resulting barycentric-corrected RVs (denoted as \texttt{CCFRVMOD} within the \texttt{NEID-DRP} documentation), from the NExScI NEID Archive\footnote{\url{https://neid.ipac.caltech.edu/}}. The RV data gathered from NEID for this study are illustrated in Figure~\ref{fig:global_model} (note that we subtract the Keplerian baseline to more clearly demonstrate the RM effect) and are available through the Data Behind the Figure program (see figure caption for link).

\section{Stellar parameters}\label{sec:stellar}

\subsection{Synthetic Spectral Fit By \texttt{iSpec}}

\par All out-of-transit NEID spectra for TOI-5126 and TOI-5398 are corrected for RV shifts, co-added, and subsequently utilized to determine the respective stellar atmospheric parameters, including stellar effective temperature (\teff), surface gravity (\logg), metallicity (\feh), and projected rotational velocity (\vsini). We use the synthetic spectral fitting technique provided by the Python package  \texttt{iSpec} \citep{Blanco2014, Blanco2019} to measure these parameters. 

We employ the SPECTRUM radiative transfer code \citep{Gray1994}, the MARCS atmosphere model \citep{gustafsson2008_MARCS}, and the sixth version of the GES atomic line list \citep{Heiter2021_GES}, all integrated within \texttt{iSpec}, to create a synthetic model for all the combined out-of-transit NEID spectra (SNR of 112 for TOI-5126; SNR of 137 for TOI-5398). In our fitting process, we treat micro-turbulent velocities as a variable, allowing us to accurately represent the small-scale turbulent motions in the stellar atmosphere. Conversely, we determine macro-turbulent velocities using an empirical relationship \citep{Doyle2014Vmac} that leverages established correlations with various stellar attributes. We select specific spectral regions to streamline the fitting process, focusing on the wings of the H$\alpha$, H$\beta$, and Mg I triplet lines, which are indicative of $\teff$ and $\logg$, as well as Fe I and Fe II lines, which are crucial for constraining $\feh$ and $\vsini$. We then utilize the Levenberg-Marquardt nonlinear least-squares fitting algorithm \citep{Markwardt2009} to refine our spectroscopic parameters by iteratively minimizing the $\chi^2$ value between the synthetic and observed spectra. The final spectroscopic parameters are detailed in Table~\ref{tab:results}.

\subsection{SED+MIST Fit By \texttt{EXOFASTv2}}

To ascertain additional stellar parameters, such as stellar mass (\mstar) and radius (\rstar), we utilize the MESA Isochrones \& Stellar Tracks (MIST) model \citep{Choi2016mist,Dotter2016mist} in combination with a spectral energy distribution (SED) fit. Photometry for the SED fit was compiled from various catalogs, including 2MASS \citep{Cutri2003}, WISE \citep{Cutri2014AllWISE}, TESS \citep{Ricker2015}, and \gaia DR3 \citep{GaiaCollaboration2023}. We apply Gaussian priors based on our synthetic spectral fit to \teff and \feh, along with the parallax from \gaia DR3, and an upper limit for the $V$-band extinction from \cite{Schlafly2011}. To accommodate the approximate 2.4\% systematic uncertainty floor in \teff, as suggested by \cite{Tayar2022}, we increase the uncertainties of \teff to 150 K for both TOI-5126 and TOI-5398.

As mentioned in \cite{Fairnington2023} and \cite{Mantovan2024a} respectively, both TOI-5126 and TOI-5398 exhibit strong stellar rotation signals, providing additional stringent constraints on their stellar ages. We adopt Equation 7 from \cite{Barnes2007} to calculate the gyrochronological ages for these stars based on the stellar rotation periods derived from our work (see Section~\ref{sec:obliq}). The resulting gyrochronological ages for TOI-5126 and TOI-5398 are $0.13 \pm 0.02$ Gyr and $0.42^{+0.07}_{-0.05}$ Gyr, respectively. Subsequently, we adopt these ages and their 3$\sigma$ uncertainties as priors for the stellar ages in our SED fit.

We perform the SED fitting using the Differential Evolution Markov Chain Monte Carlo (DEMCMC) technique, integrated within \exofasttwo \citep{Eastman2017,Eastman2019}, from which we obtain uncertainty estimates. The DEMCMC procedure was considered converged when the Gelman-Rubin diagnostic \citep[$\hat{R}$;][]{Gelman1992} fell below 1.01 and the count of independent draws surpassed 1000. Our final adopted stellar parameters are listed in Table 1; we find all derived values are consistent (within $< 2 \sigma$) with results from discovery papers (see \citealt{Fairnington2023} for TOI-5126, \citealt{Mantovan2024a} for TOI-5398). 

\begin{deluxetable*}{lccccc}
\tablecaption{Stellar and Planetary Parameters of TOI-5126 and TOI-5398
\label{tab:results}}                           
\tabletypesize{\scriptsize}                    
\tablehead{\colhead{ }                         &\colhead{Description}                                                    &\colhead{Priors$^{a}$}                                                                             &\colhead{TOI-5126$^{b}$}                                                                                                                   &\colhead{Priors$^{a}$}                                                                                                                                                  &\colhead{TOI-5398}                                                         }                                                                                                                                                                                          
\startdata                                     
\multicolumn{6}{l}{\textbf{Stellar Parameters:}}\\
\multicolumn{6}{l}{{Synthetic spectral fit:}}\\
$\teff$                                        &Effective temperature (K)                                                &-                                                                                            &$6256\pm87$                                                                                                                   &-                                                                                                                                                                 &$6072\pm89$                             \\                                                                                                                                                                                                                            
$\feh$                                         &Metallicity (dex)                                                        &-                                                                                            &$0.03\pm0.07$                                                                                                                        &-                                                                                                                                                                 &$0.01\pm0.05$                           \\                                                                                                                                                                                                                            
$\logg$                                        &Surface gravity (log$_{10}$(cm/s$^2$))                                   &-                                                                                            &$4.39\pm0.20$                                                                                                                        &-                                                                                                                                                                 &$4.63\pm0.16$                           \\                                                                                                                                                                                                                            
\multicolumn{6}{l}{{SED+MIST fit (adopted):}}\\
$M_*$                                          &Stellar mass ($M_\odot$)                                                 &-                                                                                            &$1.248^{+0.035}_{-0.034}$                                                                                                            &-                                                                                                                                                                 &$1.105\pm0.029$                         \\                                                                                                                                                                                                                  
$R_*$                                          &Stellar radius ($R_\odot$)                                               &-                                                                                            &$1.199\pm0.033$                                                                                                            &-                                                                                                                                                                 &$1.035\pm0.028$               \\                                                                                                                                                                                                                            
$\teff$                                        &Effective temperature (K)                                                &$\mathcal{N}(6256\vert 6256,150)$                                                            &$6297^{+88}_{-86}$                                                                                                                          &$\mathcal{N}(6072\vert 6072,150)$                                                                                                                                 &$6039^{+80}_{-81}$                     \\                                                                                                                                                                                                                             
$\feh$                                         &Metallicity (dex)                                                        &$\mathcal{N}(0.03\vert 0.03,0.07)$                                                             &$0.056^{+0.068}_{-0.065}$                                                                                                            &$\mathcal{N}(0.01\vert 0.01,0.05)$                                                                                                                                &$0.014^{+0.050}_{-0.049}$               \\                                                                                                                                                                                                                            
$\logg$                                        &Surface gravity (log$_{10}$(cm/s$^2$))                                   &-                                                                                            &$4.377\pm0.019$                                                                                                                      &-                                                                                                                                                                 &$4.451^{+0.020}_{-0.019}$               \\                                                                                                                                                                                                                                      
Age                                            &Age (Gyr)                                                                &$\mathcal{N}(0.13\vert 0.13,0.06)$                                                           &$0.170^{+0.059}_{-0.049}$                                                                                                            &$\mathcal{N}(0.42\vert 0.42,0.21)$                                                                                                                                &$0.54^{+0.20}_{-0.17}$                     \\                                                                                                                                                                                                                         
$\varpi$                                       &Parallax (mas)                                                           &$\mathcal{N}(6.257\vert 6.257,0.024)$                                                        &$6.256^{+0.025}_{-0.026}$                                                                                                                      &$\mathcal{N}(7.648\vert 7.648,0.017)$                                                                                                                             &$7.649\pm0.018$                         \\                                                                                                                                                                                                                            
d                                              &Distance (pc)                                                            &-                                                                                            &$159.84^{+0.66}_{-0.65}$                                                                                                                      &-                                                                                                                                                                 &$130.74^{+0.30}_{-0.31}$                         \\                                                                                                                                                                                                                            
\multirow{ 2}{*}{$v\sin i_*$ {\tiny (\texttt{iSpec})}}   &Projected stellar                                                        &\multirow{ 2}{*}{-}                                                                          &\multirow{ 2}{*}{$13.84\pm0.80$}                                                                                                     &\multirow{ 2}{*}{-}                                                                                                                                               &\multirow{ 2}{*}{$7.65\pm0.54$}         \\                                                                                                                                                                                                                            
                                               &rotational velocity (km/s)                                               &                                                                                             &                                                                                                                                     &\\                                                                                                                                                                
\hline \\                                      
\multicolumn{6}{l}{\textbf{Rossiter-McLaughlin Parameters:}}\\
$\lambda$                                      &Sky-projected spin-orbit angle (deg)                                     &$\mathcal{U}(0\vert -180,180)$                                                               & $1\pm48$                                                                                                   &$\mathcal{U}(0\vert -180,180)$                                                                                                                                    & $-8.1_{-6.3}^{+5.3}$ 		 \\                                                                                                                                                                                                                         
\multirow{ 2}{*}{$v \sin{i_\star}$}            &Projected stellar                                                        &$\multirow{ 2}{*}{$\mathcal{N}(13.84\vert 13.84,0.80)$}$                                     & \multirow{ 2}{*}{ $13.77\pm0.79$ }                                                                                                  &$\multirow{ 2}{*}{$\mathcal{N}(7.65\vert 7.65,0.54)$}$                                                                                                            & \multirow{ 2}{*}{$7.58\pm0.19$}                  \\                                                                                                                                                                                                              
                                               &rotational velocity (km/s)                                               &                                                                                             &                                                                                                  &                                                                                                                                                                  &                          \\                                                                                                                                                   
$\xi$                                          &Micro-turbulent velocity (km/s)                                          &$\mathcal{T}(1.23\vert 1.23, 1, 0, 10)$                                                      & $1.37_{-0.81}^{+0.93}$                                                                                                   &$\mathcal{T}(1.14\vert 1.14, 1, 0, 10)$                                                                                                                           & $1.20_{-0.68}^{+0.74}$ 	  \\                                                                                                                                                                                                       
$\zeta$                                        &Macro-turbulent velocity (km/s)                                          &$\mathcal{T}(4.52\vert 4.52, 1, 0, 10)$                                                      & $4.57\pm0.98$                                                                                                   &$\mathcal{T}(3.77\vert 3.77, 1, 0, 10)$                                                                                                                           & $3.63\pm0.75$  \\                                                                                                                                                                                                                 
\hline \\                                      
\multicolumn{6}{l}{\textbf{Planetary Parameters:}}\\
$R_b / R_\star$                                &Planet-to-star radius ratio                                              &$\mathcal{U}(0.0348\vert 0, 1)$                                                              & $0.03612\pm0.00042$                                                                                   &$\mathcal{U}(0.0898\vert 0, 1)$                                                                                                                                   & $0.09116\pm0.00059$                                              \\                                                                                                                                                                                             
\multirow{ 2}{*}{$(R_\star + R_b) / a_b$}      &Ratio of the sum of star and                                             &\multirow{ 2}{*}{$\mathcal{U}(0.0916\vert 0, 1)$}                                            & \multirow{ 2}{*}{ $0.08951_{-0.00078}^{+0.0015}$ }                                                                                  &\multirow{ 2}{*}{$\mathcal{U}(0.0543\vert 0, 1)$}                                                                                                                 & \multirow{ 2}{*}{ $0.05295_{-0.00080}^{+0.0010}$ }                                             \\                                                                                                                                                                                                          
                                               &planet radii to semi-major axis                                          &                                                                                              &                                                                                  &                                                                                                                                                                   &                                             \\                                                                                                                                                                                                
$\cos{i_b}$                                    &Cosine of inclination                                                    &$\mathcal{U}(0.0\vert 0, 1)$                                                                 & $0.0097_{-0.0067}^{+0.0091}$                                                                                   &$\mathcal{U}(0.0\vert 0, 1)$                                                                                                                                      & $0.0103_{-0.0046}^{+0.0040}$                                              \\                                                                                                                                                                                                 
$T_{0;b}$                                      &Mid-transit time - 2459600 ($\rm BJD_{\rm TDB}$)                         &$\mathcal{U}(27.039 \vert 26.939 , 27.139 )$                                                 & $27.03719_{-0.00043}^{+0.00053}$                                                                              &$\mathcal{U}(16.492\vert 16.392, 16.592)$                                                                                                                                    & $16.49202\pm0.00022$                                              \\                                                                                                                                                                                                       
$P_b$                                          &Orbital period (days)                                                    &$\mathcal{U}(5.458\vert 5.358, 5.558)$                                                       & $5.4588627_{-0.0000085}^{+0.0000077}$                                                                                  &$\mathcal{U}(10.591\vert 10.491, 10.691)$                                                                                                                         & $10.590535\pm0.000024$                                              \\                                                                                                                                                                                          
$K_b$                                          &Radial velocity semi-amplitude (m/s)                                     &$\mathcal{N}(6.6\vert 6.6,2.5)$                                                              &$6.6\pm2.5$                                                                                                                    &$\mathcal{N}(15.7\vert 15.7,1.5)$                                                                                                                                       & $15.2\pm1.3$ 				                            \\                                                                                                                                                                      
$q_{1;\rm NEID}$                               &Linear limb-darkening coefficient for NEID                               &$\mathcal{U}(0.5\vert 0, 1)$                                                                 &$0.51\pm0.34$                                                                                                                        &$\mathcal{U}(0.5\vert 0, 1)$                                                                                                                                      & $0.31_{-0.22}^{+0.35}$                              \\                                                                                                                                                                                                                                            
$q_{2;\rm NEID}$                               &Quadratic limb-darkening coefficient for NEID                            &$\mathcal{U}(0.5\vert 0, 1)$                                                                 &$0.46_\pm0.31$                                                                                                               &$\mathcal{U}(0.5\vert 0, 1)$                                                                                                                                               & $0.17_{-0.12}^{+0.33}$                                        \\                                                                                                                                                                                                                                                     
$q_{1;\rm HARPS-N}$                               &Linear limb-darkening coefficient for HARPS-N                               &-                                                                 &-                                                                                                                        &$\mathcal{U}(0.5\vert 0, 1)$                                                                                                                                                                       & $0.873_{-0.15}^{+0.089}$                              \\                                                                                                                                                                                                                                            
$q_{2;\rm HARPS-N}$                               &Quadratic limb-darkening coefficient for HARPS-N                            &-                                                                 &-                                                                                                               &$\mathcal{U}(0.5\vert 0, 1)$                                                                                                                                                                                & $0.381\pm0.091$                                       \\
\hline \\                                      
\multicolumn{6}{l}{\textbf{Derived Parameters:}}\\
$M_b$                                          &Planetary mass ($\mathrm{M_{J}}$)                                      &-                                                                                              & $0.067^{+0.028}_{-0.022}$                                                                                                            &-                                                                                                                                                                   & $0.135\pm0.043$                   \\                                                                                                                                                                                                    
$R_b$                                          &Planetary radius ($\mathrm{R_{J}}$)                                    &-                                                                                              & $0.421\pm0.014$                                                                                                            &-                                                                                                                                                                   & $0.921\pm0.026$         \\                                                                                                                                                                                                              
$a_b / R_\star$                                &Semi-major axis scaled by stellar radius                                 &-                                                                                            & $11.57_{-0.18}^{+0.10}$                                                                                                            &-                                                                                                                                                                 & $20.61_{-0.39}^{+0.31}$                        \\                                                                                                                                                                                                               
$a_b$                                          &Semi-major axis (AU)                                                     &-                                                                                            & $0.0644\pm0.0020$                                                                                                            &-                                                                                                                                                                 & $0.0994\pm0.0032$               \\                                                                                                                                                                                                            
$i_b$                                          &Inclination (deg)                                                        &-                                                                                            & $89.44_{-0.51}^{+0.39}$                                                                                                            &-                                                                                                                                                                 & $89.41_{-0.23}^{+0.26}$                                 \\                                                                                                                     
$T_{14;b}$                                     &Total transit duration (hours)                                           &-                                                                                            & $3.711_{-0.020}^{+0.022}$                                                                                                             &-                                                                                                                                                                 & $4.205\pm0.017$                                    \\                                                         
$u_{1;\rm NEID}$                               &Linear limb-darkening coefficient for NEID                               &-                                                                                            &$0.56_{-0.38}^{+0.57}$                                                                                                               &-                                                                                                                                                                 & $0.162_{-0.099}^{+0.22}$ 										 \\                                                                                                                                                                                                                                            
$u_{2;\rm NEID}$                               &Quadratic limb-darkening coefficient for NEID                            &-                                                                                            &$0.04\pm0.42$                                                                                                                        &-                                                                                                                                                                 & $0.32_{-0.32}^{+0.36}$ 									 \\      
$u_{1;\rm HARPS-N}$                               &Linear limb-darkening coefficient for HARPS-N                               &-                                                                                            &-                                                                                                               &-                                                                                                                                                                                & $0.70\pm0.17$ 										 \\                                                                                                                                                                                                                                            
$u_{2;\rm HARPS-N}$                               &Quadratic limb-darkening coefficient for HARPS-N                            &-                                                                                            &-                                                                                                                        &-                                                                                                                                                                       & $0.22\pm0.17$									 \\ 
\multicolumn{6}{l}{\textbf{GP Parameters:}}\\
$ln\rm B_{NEID}$& Amplitude scaling factor&-&-&$\mathcal{U}(0\vert -10, 10)$&$-6.65_{-0.63}^{+0.53}$\\ 
$ln\rm C_{NEID}$ & Balance factor&-&-&$\mathcal{U}(0\vert -10, 10)$&$-1.0_{-7.2}^{+8.2}$\\
$ln\rm L_{NEID}$ & Coherence timescale&-&-&$\mathcal{U}(0\vert -10, 10)$&$-1.0_{-7.1}^{+4.6}$\\
$ln\rm P_{rot, NEID}$& Stellar rotation period&-&-&$\mathcal{N}(2\vert 2, 0.2)$&$2.02\pm0.18$\\ 
$ln\rm B_{HARPS-N}$ & Amplitude scaling factor&-&-&$\mathcal{U}(0\vert -10, 10)$& $-7.17\pm0.15$\\ 
$ln\rm C_{HARPS-N}$ & Balance factor&-&-&$\mathcal{U}(0\vert -10, 10)$&$-0.9_{-7.3}^{+8.3}$\\
$ln\rm L_{HARPS-N}$ & Coherence timescale&-&-&$\mathcal{U}(0\vert -10, 10)$&$-1.15_{-0.96}^{+0.62}$\\
$ln\rm P_{rot, HARPS-N}$ & Stellar rotation period&-&-&$\mathcal{N}(2\vert 2, 0.2)$&$2.01\pm0.18$\\                  
\enddata                                  
\tablenotetext{a}{The notation $\mathcal{U}(p_0 | a, b)$, $\mathcal{N}(p_0 | \mu, \sigma)$, and $\mathcal{T}(p_0 | \mu, \sigma, a, b)$ represent the uniform, Gaussian, and truncated Gaussian distributions, respectively. Here, $p_0$ denotes the initial guess, $a$ and $b$ the lower and upper bounds, $\mu$ the median value, and $\sigma$ the standard deviation.}
\tablenotetext{b}{For TOI-5126\,b, we adopt the results of \cite{Fairnington2023} as priors for our fitted RV semi-amplitude $K_b$ and we directly adopt their planetary mass estimate $M_b$, which was obtained via the \cite{Otegi2020} mass-radius relation}.
\end{deluxetable*}

\section{Obliquity Modeling} \label{sec:obliq}
We determine the sky-projected spin-orbit angle $\lambda$ for TOI-5126\,b and TOI-5398\,b by jointly fitting transit photometry from their respective discovery papers (including two-minute cadence TESS light curves), and all in-transit and out-of-transit RVs from our NEID observations (see Section \ref{sec:obs}), using a modified version of the \texttt{allesfitter} package \citep{Gunther2021}. The modified version of \texttt{allesfitter} incorporates the \cite{Hirano2011} RM model implemented in \texttt{tracit} \citep{Hjorth2021, KnudstrupAlbrecht2022}. We exclude the transit data for the non-sub-Saturn planet in both systems but model their RVs (TOI-5126 c at $P=17.9 \, \mathrm{d}$ and TOI-5398 c at $P=4.77 \, \mathrm{d}$). 

For TOI-5126\,b, we adopt the Presearch Data Conditioning Simple Aperture Photometry (PDCSAP) light curves from the TESS Science Processing Operations Center (SPOC), modeling eight full transits and one partial transit from Sector 45, four full transits and one partial transit from Sector 46, and four full transits from Sector 48. Within \texttt{allesfitter}, we simultaneously fit the TESS and NEID data together with all photometric data utilized in TOI-5126\,b's discovery paper \citep{Fairnington2023}, including observations from the \emph{CHaracterising} \emph{ExOplanets} \emph{Satellite} (CHEOPS, \citealt{Benz2021}) as well as ground-based $1.0\,\mathrm{m}$ telescopes at McDonald Observatory in Fort Davis, TX and at the Cerro Tololo Inter-American Observatory (CTIO) in Cerro Tololo, Chile, which are both part of the Las Cumbres Observatory Global Network (LCOGT; \citealt{Brown2013}). Due to their large scatter relative to the expected planetary RV semi-amplitudes, we do not make use of the out-of-transit RVs presented in \cite{Fairnington2023}, which were obtained from the Tillinghast Reflector Echelle Spectrograph (TRES) at the Fred Lawrence Whipple Observatory (FLWO) and the CHIRON spectrograph on the SMARTS telescope at CTIO.

For TOI-5398\,b, we also consider the PDCSAP TESS light curves (we did not find substantive evidence for the over-corrections in the PDCSAP data reported by \citealt{Mantovan2024a}), modeling two full transits from TESS Sector 48, as well as photometry from the LCOGT $1.0\,\mathrm{m}$ telescopes at McDonald Observatory and Teide Observatory and from the $1.2 \, \mathrm{m}$ KeplerCam at FLWO on Mount Hopkins, Arizona (all sourced from \citealt{Mantovan2024a}). We additionally incorporate the out-of-transit RVs from \cite{Mantovan2024a}, which were obtained from the High Accuracy Radial velocity Planet Searcher (HARPS-N, \citealt{Cosentino2012}) at the Italian Telescopio Nazionale Galileo (TNG). Recently, \cite{Mantovan2024b} published an independent RM effect measurement of TOI-5398\,b; therefore, we model the RM effect for this target both with and without the inclusion of their published in-transit RVs (holding all other factors equal), which were also obtained from the HARPS-N spectrograph\footnote{Our work was otherwise prepared concurrently and independently to that of Mantovan et al.}. To account for strong stellar activity in the RVs, we apply a rotational Gaussian Process (GP) regression kernel formulated by \cite{DFM2017RotationGP},
\begin{eqnarray}
k(\tau) & = & \frac{B}{2+C} e^{-\tau / L}\left[\cos \left(\frac{2 \pi \tau}{P_{\mathrm{rot}}}\right)+(1+C)\right]
\end{eqnarray}
where $P_{\mathrm{rot}}$ represents the period of stellar rotation, $L$ denotes the coherence timescale, $\tau$ signifies the interval between two successive data points, $B$ is a scaling hyperparameter that adjusts the amplitude, and $C$ serves as the balance parameter for the periodic and nonperiodic components of the GP kernel. For TOI-5398, we note that our adopted GP model is most similar to ``Case 1'' from \cite{Mantovan2024a}, though the results from their multi-dimensional ``Case 3'' were ultimately adopted.

For both TOI-5126\,b and TOI-5398\,b, we adopt priors but utilize the results reported in the literature (\citealt{Fairnington2023} and \citealt{Mantovan2024a}, respectively) as our initial guesses for $R_\mathrm{p}/R_*$, $(R_*+R_\mathrm{p})/a$, $\cos(i)$, $T_0$, $P$, and $K$ (see Table \ref{tab:results} for parameter definitions). We fixed eccentricity ($e$) and the argument of periastron ($\omega$) to 0 due to the near-zero eccentricity reported by the discovery papers of TOI-5126\,b and TOI-5398\,b. We consider transformed linear and quadratic limb-darkening coefficients $q_1$ and $q_2$, in addition to physical linear and quadratic limb-darkening coefficients $u_1$ and $u_2$\footnote{The relationship between the transformed ($q_1$, $q_2$) and physical ($u_1$, $u_2$) limb-darkening coefficients is described in \cite{Kipping2013} by Equation (15): $u_1=2\sqrt{q_1}q_2$ and Equation (16): $u_2=\sqrt{q_1}(1-2q_2)$}, for both NEID and HARPS-N, initializing $q_1$ and $q_2$ at 0.5 for each instrument. As an initial guess for \vsini, for which we apply Gaussian priors, we adopt the values derived from our synthetic spectral fit. Finally, we initialize $\lambda$ at $0\degree$ for both systems. All parameters (except $e$ and $\omega$) are allowed to vary during the fitting process. To account for the short-term overnight instrumental systematics and stellar variability in the RM fit, we employ a quadratic baseline for TOI-5126 and a constant baseline for TOI-5398, though we note that all tested baselines (constant, linear, quadratic, and cubic) produced consistent $\lambda$ values (within $1\sigma$).

For the fits of both systems, we sample the posterior distributions for all modeled parameters using an affine-invariant Markov Chain Monte Carlo (MCMC) grounded in \texttt{emcee} \citep{emcee} with 100 walkers. Best-fit parameters were obtained after 200,000 accepted steps per walker were reached (note that each ran 10,000 burner steps); our final MCMC results and associated $1 \sigma$ uncertainties are summarized in Table \ref{tab:results}. Additionally, all Markov Chains reached $>50 \times$ their autocorrelation lengths, indicating convergence.

\begin{figure*}[!ht]
    \centering
    \includegraphics[width=1\textwidth]{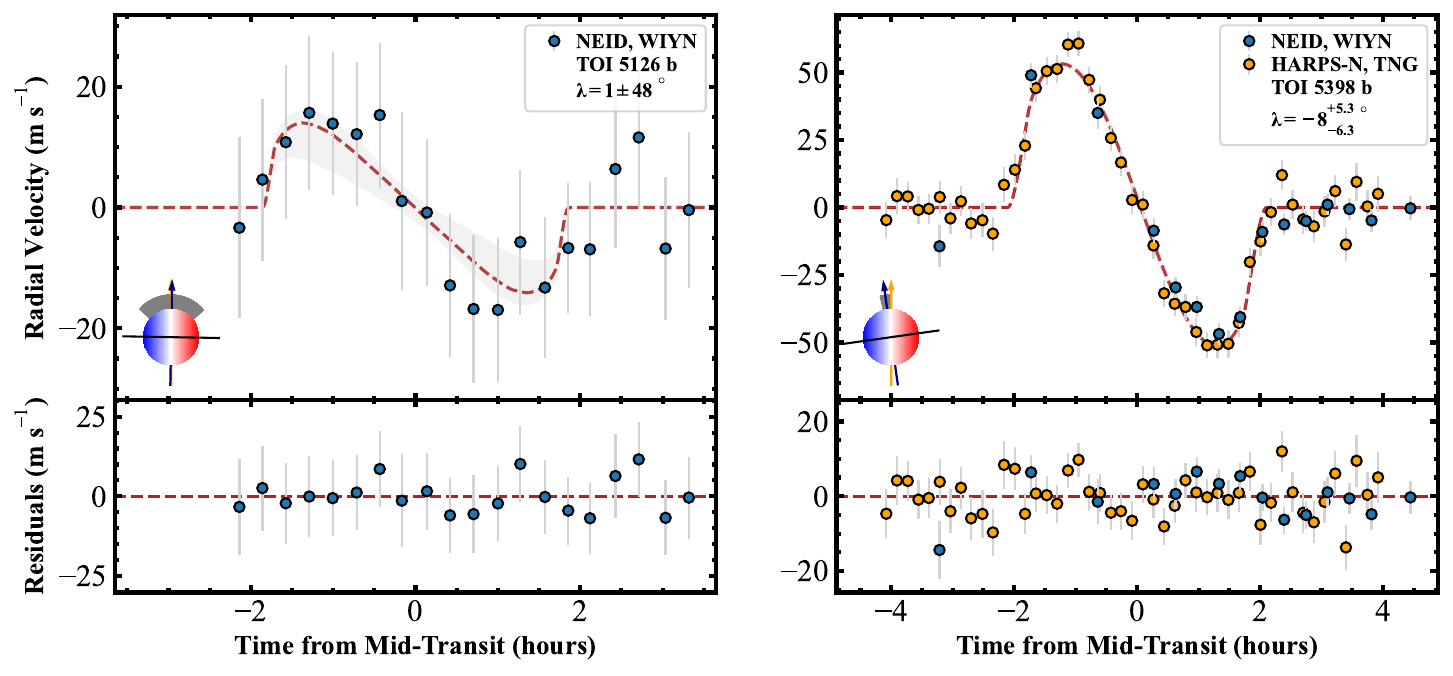}
    \caption{Keplerian signal-subtracted radial velocities (top panels) and RM model residuals (bottom panels) for TOI-5126 (left) and TOI-5398 (right). In the top panels, the best-fit RM models are shown with red dashed lines, and their uncertainties are indicated by grey shadows, while the planets' spin-orbit configurations are depicted in the bottom left portion of the plot. Note that the NEID RV measurement obtained $\approx1$ hour prior to ingress for TOI 5398\,b was taken on April 10, 2022 via 300-second exposure as part of our out-of-transit follow-up for this system. All NEID data used in this work are available \href{https://github.com/BrandonRadzom/stellar-obliquities.git}{here}.} 
    \label{fig:global_model}
\end{figure*}

We find good agreement ($< 2 \sigma$ difference) with all stellar, planetary, and derived parameters reported in the discovery papers of both systems (\citealt{Fairnington2023} for TOI-5126\,b and \citealt{Mantovan2024a} for TOI-5398\,b), except $T_{0;b}$ (2.5$\sigma$ discrepant) for TOI-5126, which may be due to tentative transit timing variations (see \citealt{Fairnington2023}). Critically, our RM fits show that both sub-Saturns are consistent with spin-orbit alignment. For TOI-5126\,b, we find a best-fit projected obliquity of $\lambda=1\pm48 \degree$. For TOI-5398\,b, we compute a best-fit $\lambda=-8.1_{-6.3}^{+5.3} \degree$ using in-transit RVs from both NEID (presented in this work) and HARPS-N (presented in \citealt{Mantovan2024b}), while we derive a marginally aligned value of $\lambda=-24_{-13}^{+14} \degree$ using our NEID data alone (note that both values are within $2\sigma$ of the $\lambda=3.0_{-4.2}^{+6.8}$ reported by \citealt{Mantovan2024b}). Figure \ref{fig:global_model} reveals the best-fit joint models and associated residuals for both systems, while Table \ref{tab:results} contains all fitted and derived parameters.

It is useful to additionally measure the 3D stellar obliquity $\psi$, which yields a system's true spin-orbit configuration.
To estimate $\psi$, we first derive the stellar inclination $i_*$ and stellar equatorial velocity $v_{eq}$ for both TOI-5126 and TOI-5398 by applying the Bayesian inference method presented by \cite{Masuda2020stincl} and \cite{Hjorth2021} to our values for stellar radius $\rstar$, stellar rotation period $\prot$, and $\cos i_{*}$.
We utilize the aforementioned TESS PDCSAP light curves to compute the stellar rotation periods for each target, employing the autocorrelation function (ACF) implemented in \texttt{SpinSpotter} \citep{Holcomb2022spinspotter}. Correspondingly, we find $P_{\rm rot}=4.05 \pm 0.21$ days for TOI-5126 and $P_{\rm rot}=7.443 \pm 0.041$ days for TOI-5398. However, due to the effects of latitudinal differential rotation \citep{Epstein2014, Aigrain2015} and the systematic uncertainty floor (of $\approx$\,4.2\%) on the derivation of stellar radii \citep{Tayar2022}, we adopt a 10\% uncertainty floor on our derived stellar rotation periods, yielding $P_{\rm rot}=4.05 \pm 0.41$ days and $P_{\rm rot}=7.443 \pm 0.74$ days for TOI-5126 and TOI-5398, respectively. These estimates are consistent with those reported by their respective discovery papers, which both applied the Lomb-Scargle periodogram to photometric data, rather than the ACF, in order to ascertain their adopted rotation periods (\citealt{Fairnington2023} find $P_{\mathrm{rot}} = 4.602^{+0.071}_{-0.067}$ days for TOI-5126, \citealt{Mantovan2024a} find $P_{\mathrm{rot}} = 7.34 \pm 0.05$ days for TOI-5398). From our Bayesian analysis, we find $i_*=90 \pm 15\degree$ and $v_{eq} = 14.36 \pm 1.02 \, \mathrm{km/s}$ for TOI-5126 and $i_*=91 \pm 17\degree$ and $v_{eq} = 8.09 \pm 0.87 \, \mathrm{km/s}$ for TOI-5398, and compute the true 3D stellar obliquities as (see Eqn. 9 of \citealt{Fabrycky2009}):
\begin{equation}
    \cos{\psi} = \cos{i_*} \cos{i} + \sin{i_*} \sin{i} \cos{\lambda}
\end{equation}
where $i$ is the inclination angle of the planet. For TOI-5126\,b, we find a true obliquity of $\psi=37.1^{+20.3}_{-33.6} \degree$. Similarly, we derive $\psi=16.4^{+7.6}_{-10.3}\degree$ for TOI-5398\,b (in good agreement with the $\psi=13.2\pm8.2 \degree$ derived by \citealt{Mantovan2024b}), demonstrating that both systems are consistent with alignment.

\section{Discussion} \label{sec:discussion}
\begin{figure*}[htb]
    \centering
    \includegraphics[width=1\textwidth,trim=0 20 0 20,clip]{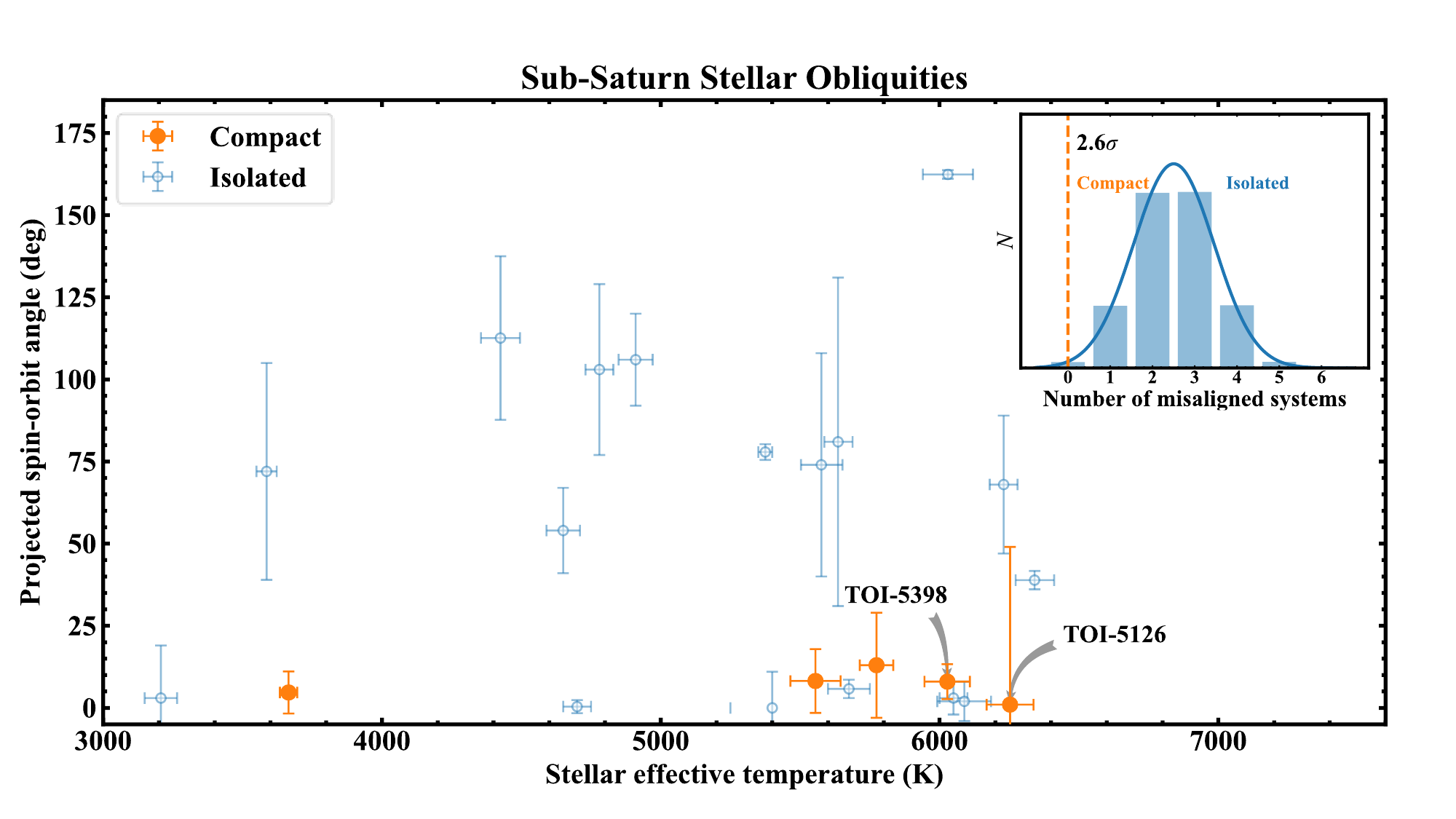}
    \caption{Sky-projected stellar obliquity $\lambda$ versus stellar effective temperature \teff for all sub-Saturns considered in this work, with $1\sigma$ error bars shown. The sub-panel in the top right displays the resultant distributions of the number of misaligned sub-Saturns found in compact multi-planet systems (dashed line) and the number of misaligned isolated sub-Saturns after performing 100,000 random draws (bars), with statistical significance ($2.6\sigma$) indicated. In both panels, orange corresponds to sub-Saturns in compact multi-planet systems while blue corresponds to isolated sub-Saturns.} 
    \label{fig:lambda_teff}
\end{figure*}

While sub-Saturns broadly appear to boast a wide range of stellar obliquities, the few previously confirmed to be in compact multi-planet systems are exclusively aligned. Our RM measurements for two of such recently-confirmed sub-Saturns, TOI-5126\,b and TOI-5398\,b, continue this trend, as both are consistent with alignment. We show the updated stellar obliquity distribution for the sub-Saturn population as a function of stellar effective temperature in Figure \ref{fig:lambda_teff}. To construct this sub-Saturn sample, we combine the catalogs of \cite{Albrecht2022} and TEPCat\footnote{\url{https://www.astro.keele.ac.uk/jkt/tepcat/obliquity.html}} \citep{Southworth2011} accessed on March 7, 2024. We consider only RM effect measurements, which exhibit near-uniform sensitivity to the full range of possible $\lambda$ values (compared to nonuniform methods such as starspot tracking or gravity darkening, \citealt{Siegel2023, DongFM2023}) and constitute the majority of obliquity measurements \citep{Triaud2018, Albrecht2022}, prioritizing those selected in \cite{Albrecht2022} (we otherwise adopt the ``preferred'' measurements from TEPCat). Finally, we apply a mass cut of $17\, \mearth <M_{pl}<95\,\mearth$, and select only single-star systems by cross-matching our sample with the multi-star catalog of \cite{Rice2024}, which is based on the most recent data from \gaia DR3. This yields 22 sub-Saturn systems with $\lambda$ measurements to comprise our statistical sample. We further distinguish between sub-Saturns that are isolated and those in compact multi-planet systems, which we define as having a nearby companion with a period ratio of $<4$. All sub-Saturns considered in this work are presented in Table \ref{tab:fullerr_sample} (note that we supplement any missing parameters using the NASA Exoplanet Archive\footnote{\url{https://exoplanetarchive.ipac.caltech.edu}} and values adopted by their respective $\lambda$ reference papers). Our RM measurements allow for more reliable statistical analyses to be performed on the relationship between compact configurations and spin-orbit alignment for sub-Saturns (Section \ref{sec:discussion-significance}) and thus more robust constraints on the origins of spin-orbit misalignment as well as the formation histories of the sub-Saturn population (Sections \ref{sec:discussion-implications} and \ref{sec:discussion-teff-lambda}).

\begin{deluxetable*}{cccccccc}
\tabletypesize{\scriptsize}
\tablecaption{Sub-Saturns Around Single Stars With a Stellar Obliquity Measurement}\label{tab:fullerr_sample}
\tablehead{
\colhead{System}     & \colhead{Planet}                 & \colhead{$T_\mathrm{eff}$ (K)}    & \colhead{$M_*$ [$\mathrm{M}_\odot$]}  &\colhead{$P$ (d)} & \colhead{$M_{pl}$ ($\mathrm{M}_\oplus$)}  &\colhead{$\lambda$ ($\degree$)}            & \colhead{$\lambda$ Reference}}

\tablewidth{300pt}
\startdata
\multicolumn{8}{l}{\textbf{Compact Multi-planet Systems:}}\\\\ AU Mic &         b &            3665$\pm 100$ &                   0.5$\pm 0.03$ &                      8.46$\pm 0.0000052$ &  20.12$^ {+ 1.57 }_{- 1.72 } $ &    $-4.7^ {+ 6.4 }_{- 6.8 } $ &  \cite{Hirano2020b} \\
       Kepler-9 &         b &             5774$\pm 60$ &     1.02$^ {+ 0.03 }_{- 0.04 } $ &                     19.24$\pm 0.00006$ &    43.41$^ {+ 1.6 }_{- 2.0 } $ &               13$\pm 16$ &     \cite{Wang2018} \\
       WASP-148 &         b &             5555$\pm 21$ &     0.95$^ {+ 0.03 }_{- 0.05 } $ &                       8.80$\pm 0.000043$ &  91.24$^ {+ 6.99 }_{- 5.09 } $ &     $-8.2^ {+ 9.7 }_{- 8.7 } $ &     \cite{Wang2022} \\
       TOI-5126 &         b &             6297$^{+88}_{-86}$ &   1.248$^ {+ 0.035 }_{- 0.034 } $ &                      5.46$\pm 0.0000062$ &   21$^{+9}_{-7} $ &   1$\pm 48$  &              this work \\
       TOI-5398 &         b &  6039$^ {+ 80 }_{- 81 } $ &  1.105$\pm0.029 $ &  10.59$^ {+ 0.000041 }_{- 0.000046 } $ &    42.9$\pm 13.7$ &  $-8.1^ {+ 5.3 }_{- 6.3 } $ &              this work \\\\
       \hline
{\textbf{Isolated Systems:}}\\\\
GJ 436 &         b &             3586$\pm 54$ &               0.47$\pm 0.07$ &                     2.64$\pm 0.00000057$ &                 22.25$\pm 2.23$ &   72$^ {+ 33 }_{- 24 } $ &          \cite{Bourrier2018} \\
       HAT-P-11 &         b &             4780$\pm 50$ &  0.81$^ {+ 0.02 }_{- 0.03 } $ &                4.89$\pm 0.0000068$ &                  26.7$\pm 2.23$ &  103$^ {+ 26 }_{- 10} $ &            \cite{Winn2010b} \\
       HAT-P-12 &         b &             4650$\pm 45$ &               0.73$\pm 0.02$ &                     3.21$\pm 0.0000021$ &                 67.08$\pm 3.81$ &   $-54^ {+ 41 }_{- 13 } $ &          \cite{Mancini2018} \\
       HD 89345 &         b &  5576$^ {+ 70 }_{- 71 } $ &  1.16$^ {+ 0.04 }_{- 0.05 } $ &                    11.81$\pm 0.0002$ &     34.97$^ {+ 5.4 }_{- 5.72 } $ &   74.2$^ {+ 33.6 }_{- 32.5 } $ &          \cite{Bourrier2023}\\
      HD 332231 &         b &  6089$^ {+ 97 }_{- 96 } $ &               1.13$\pm 0.08$ &                    18.71$\pm 0.00043$ &                 77.57$\pm 6.68$ &                  $-2\pm 6$ &     \cite{KnudstrupAlbrecht2022}  \\
      HIP 67522 &         b &             5675$\pm 75$ &               1.22$\pm 0.05$ &  6.96$^ {+ 0.000016 }_{- 0.000015 } $ &      $79^{a} $ &      $-5.8^ {+ 2.8 }_{- 5.7 } $ &         \cite{Heitzmann2021} \\
          K2-25 &         b &             3207$\pm 58$ &               0.26$\pm 0.01$ &  3.48$^ {+ 0.0000006 }_{- 0.0000005 } $ &      24.51$^ {+ 5.7 }_{- 5.2 } $ &                 3$\pm 16$ &        \cite{Stefansson2020}  \\
         K2-105 &         b &             5636$\pm 70$ &               1.05$\pm 0.02$ &  8.27$^ {+ 0.000007 }_{- 0.0000067 } $ &                 30.01$\pm 19.0$ &   $-81^ {+ 50}_{- 47 } $ &          \cite{Bourrier2023}  \\
        KELT-11 &         b &             5375$\pm 25$ &               1.44$\pm 0.07$ &                     4.74$\pm 0.00003$ &                 54.36$\pm 4.77$ &     $-77.86^ {+ 2.36 }_{- 2.26 } $ &           \cite{Mounzer2022} \\
       TOI-1842 &         b &             6230$\pm 50$ &               1.46$\pm 0.03$ &  9.57$^ {+ 0.0002 }_{- 0.0001 } $ &  68.03$^ {+ 12.72 }_{- 12.08 } $ &   $-68.1^ {+ 21.2 }_{- 14.7 } $ &        \cite{Hixenbaugh2023} \\
       TOI-1859 &         b &  6341$^ {+ 68 }_{- 70 } $ &               1.29$\pm 0.06$ &                    63.48$\pm 0.0001$ &      $68.67^{a}$ &     38.9$^ {+ 2.8 }_{- 2.7 } $ &              \cite{Dong2023} \\
        WASP-39 &         b &             5400$\pm 50$ &   0.93$\pm 0.03 $ &                     4.06$\pm 0.000009$ &                 89.01$\pm 9.54$ &                 0$\pm 11$ &           \cite{Mancini2018} \\
        WASP-69 &         b &             4700$\pm 50$ &               0.98$\pm 0.14$ &                     3.87$\pm 0.000002$ &                 92.19$\pm 9.54$ &      0.4$^ {+ 2.0 }_{- 1.9 } $ &  \cite{Casasayas-Barris2017} \\
       WASP-107 &         b &             4425$\pm 70$ &               0.68$\pm 0.02$ &                5.72$\pm 0.000002$ &                 30.52$\pm 1.59$ &  118.1$^ {+ 37.8 }_{- 19.1 } $ &         \cite{Rubenzahl2021} \\
       WASP-131 &         b &             6030$\pm 50$ &               1.06$\pm 0.06$ &                     5.32$\pm 0.000005$ &                 85.83$\pm 6.36$ &    162.4$^ {+ 1.3 }_{- 1.2 } $ &             \cite{Doyle2023} \\
       WASP-156 &         b &             4910$\pm 61$ &               0.84$\pm 0.05$ &                     3.84$\pm 0.000003$ &    40.69$^ {+ 3.18 }_{- 2.86 } $ &               $105.7^{+14.0}_{-14.4}$ &          \cite{Bourrier2023}  \\
       WASP-166 &         b &             6050$\pm 50$ &               1.19$\pm 0.06$ &                     5.44$\pm 0.000004$ &                 32.11$\pm 1.59$ &                  3$\pm 5$ &           \cite{Hellier2019}
\enddata
\tablenotetext{a}{Calculated using the \cite{ChenKipping2017} mass-radius relationship.}
\end{deluxetable*} 
   
\subsection{Sub-Saturns in Compact Multi-planet Systems are Significantly Aligned}
\label{sec:discussion-significance}
It is clear from Figure \ref{fig:lambda_teff} that spin-orbit alignment is much more common for sub-Saturns in compact multi-planet systems than for isolated sub-Saturns, while no clear relation exists for the host star's effective temperature (as seen in previous works, e.g., \citealt{Albrecht2022}). Our two stellar obliquity measurements increase the sample of sub-Saturns in compact configurations to five, enabling the first statistical tests of this trend. We investigate the statistical significance of preferential alignment in compact systems by comparing the number of misaligned systems found in the compact multi-planet sub-Saturn sample (zero) with that found in a series of random draws from the isolated sub-Saturn sample. We perform this statistical analysis using the projected 2D spin-orbit angles $\lambda$, rather than the true 3D spin-orbit angles $\psi$, for two reasons: i) considering solely $\psi$ is statistically prohibitive, 
as only 14 of the 22 systems in our sample have $\psi$ measurements, and ii) nearly all 22 systems in our sample orbit cooler stars ($\teff<6250\,\mathrm{K}$), which tend to have edge-on inclinations such that $\psi\approx|\lambda|$ (see \citealt{Louden2021}; note that the average difference between $\psi$ and $|\lambda|$ is just $13.2\degree$ for the 14 systems with measurements for both angles).

Of the 22 sub-Saturns with $\lambda$ measurements considered in this work, five are in compact multi-planet systems, while the other 17 are isolated. Of the five in compact configurations, none are misaligned, while 9/17 (53\%) isolated sub-Saturns are misaligned. We define misalignment as $|\lambda|>10 \degree$ at the $1\sigma$ level (i.e., $|\lambda|-\sigma_{\lambda}>10\degree$) and $\lambda>0 \degree$ at the $2\sigma$ level (i.e., $|\lambda|-2\sigma_{\lambda}>0\degree$), where we adopt the $1\sigma$ uncertainties ($\sigma_\lambda$) reported for each $\lambda$ measurement (see Table \ref{tab:fullerr_sample}). In order to perform a fair comparison between the compact multi-planet sample and the isolated sample, we randomly select and classify the alignment of five isolated sub-Saturns. We perform 100,000 iterations of these random draws, counting the number of misaligned systems for each, and fit a Gaussian function over the resulting distribution of $N$ systems per count. The upper-right sub-panel of Figure \ref{fig:lambda_teff} reveals the result, illustrating that sub-Saturns in compact multi-planet systems are more often aligned than isolated sub-Saturns to a significance level of $2.6 \sigma$.

We verified that this significance is robust to the precise misalignment criteria adopted by re-running the above procedure with the following alternative cuts: $\lambda > 15 \degree$ and $\lambda >0 \degree$ at the $2\sigma$ level ($2.6\sigma$ significance), (ii) $\lambda > 20 \degree$ and $\lambda >0 \degree$ at the $2\sigma$ level ($2.6\sigma$ significance), (iii) $\lambda > 15 \degree$ ($2.3\sigma$ significance), and (iv) $\lambda >20 \degree$ ($2.3\sigma$ significance). Assuming our original alignment criteria, we additionally vary the Neptune/sub-Saturn mass boundary and re-run our procedure: (i) $15 \, \mathrm{M}_\oplus$ ($3.0 \sigma$ significance), and (ii) $0.1\,\mathrm{M}_\mathrm{J}$ ($1.8 \sigma$ significance). We note that while the significance appears to degrade with an increasing mass bound, this trend is likely not astrophysical, as the significance is directly proportional to the number of sub-Saturns in compact systems considered, which drops from six assuming a $15 \, \mathrm{M}_\oplus$ bound to only three assuming a $0.1\, \mathrm{M}_\mathrm{J}$ ($\approx 32 \, \mathrm{M}_\oplus$) bound. Finally, we note that the young ($\sim 20 \, \mathrm{Myr}$), aligned, compact sub-Saturn system AU Mic still hosts a dusty debris disk \citep{Plavchan2020,Hirano2020b}, meaning that its spin-orbit angle and compactness has likely not yet had the opportunity to be affected by various pathways to trigger post-disk eccentric migration \citep{Wu2023}. Therefore, while this system provides strong support for primordial alignment, it may bias our statistical test; we re-run our original procedure excluding AU Mic (such that there are only four compact sub-Saturn systems) and find that the significance is maintained, but reduced to $2.2 \sigma$.

\subsection{Evidence for Primordial Alignment and Misalignment Via Eccentric Migration}
\label{sec:discussion-implications}

In this work, we have demonstrated that sub-Saturns in compact planetary systems are significantly more aligned than those without nearby planetary companions (at the $2.6\sigma$ level). This trend of spin-orbit alignment in compact multi-planet systems around single stars is not limited to sub-Saturns, however. Short-period ($P\lesssim 100 \, \mathrm{d}$) Jupiters provided the first robust evidence for this phenomenon. Specifically, all measured spin-orbit angles for short-period Jupiters with nearby companions, even hot Jupiters, are consistent with alignment, e.g., Kepler-30 \citep{Sanchis-Ojeda2012}, Kepler-89 \citep{Albrecht2013}, WASP-47 \citep{Sanchis-Ojeda2015}, TOI-1478 \cite{Rice2022b}, TOI-2202 \citep{Rice2023b}, TOI-1670 \citep{Lubin2023}. Conversely, isolated short-period Jupiters display a wide range of obliquities and are frequently misaligned (\citealt{Albrecht2012, Albrecht2022}). While there is a paucity of spin-orbit measurements for sub-Neptunes and super-Earths, most of these lower-mass planets reside in compact multi-planet systems, and those with secure obliquity measurements are aligned, e.g., Kepler-50, Kepler-65 \citep{Chaplin2013}, Kepler-25 \citep{Albrecht2013}, HD 106315 \citep{Zhou2018, Bourrier2023}, TRAPPIST-1 \citep{Hirano2020a}, TOI-1726 \citep{Dai2020}, HD 63433 \citep{Mann2020}, TOI-1136 \citep{Dai2023}, and TOI-2076 \citep{Frazier2023}.

As the spin-orbit angles of compact multi-planet systems should remain largely unaltered following the dispersal of the disk, the low stellar obliquities observed for sub-Saturns and other types of exoplanets in compact systems suggest that most planetary systems are initially formed spin-orbit aligned. Consequently, the dominant mechanism driving misalignment in single-star systems is likely not a universal process that operates indiscriminately across different types of systems, but instead is inherent to those with certain post-disk dynamical histories. More broadly, support for misalignment being acquired well into the post-disk phase rather than near the onset of system formation comes from evidence that i) both young stellar systems ($\lesssim100 \, \mathrm{Myr}$ old) and planets still embedded within their debris disks are aligned \citep{Kraus2020, Plavchan2020, Hirano2020b, Martioli2020, Palle2020, Albrecht2022, Johnson2022}, and ii) hot Jupiter systems that are misaligned tend to be older \citep{Hamer2022}. In combination with the large spin-orbit angles commonly observed for misaligned isolated hot Jupiters and sub-Saturns, these facts are generally consistent with violent eccentric migration as the dominant driver of misalignment \citep{Dawson2018, Wu2023}; identifying the most relevant migration pathway(s), however, is beyond the scope of this work.

\subsection{On the Temperature-Obliquity Relation} \label{sec:discussion-teff-lambda}
There is an interesting discrepancy between the obliquity distributions of isolated hot Jupiters and isolated hot sub-Saturns: hot Jupiters are commonly misaligned around hot stars ($\teff\gtrsim 6250\,\mathrm{K}$) but exclusively aligned around cool stars (i.e., the $T_{\rm eff}-\lambda$ relation, see \citealt{Winn2010a, Albrecht2012, Albrecht2022}), while hot sub-Saturns are commonly misaligned around cool stars (and no data exist for those orbiting hot stars, see Figure \ref{fig:lambda_teff}). Realignment via tidal dissipation offers a plausible explanation for this phenomenon; cool stars are susceptible to realignment by massive hot Jupiters, while lower-mass hot sub-Saturns are less likely to realign them on timescales shorter than the system's age (e.g., see Eqn. 2 of \citealt{Albrecht2012}). Hence, both hot Jupiters and hot sub-Saturns may indeed become misaligned through high-eccentricity migration, but only cool stars with hot Jupiters can be realigned. This tidal realignment mechanism, however, requires finely-tuned parameters in order to completely realign all hot Jupiters around cool stars without leaving any on polar or retrograde orbits \citep{Rogers2013,Li2016}, which are not seen in observations.

Alternatively or in addition to this tidal realignment scenario, it is possible that sub-Saturns are subject to misalignment mechanisms that may not operate for Jupiters. For instance, \cite{Petrovich2020} illustrates that secular resonance between an outer gas giant companion and an inner sub-Saturn, both embedded in a decaying disk, can drive the sub-Saturn to a polar orbit. The level of obliquity excitation depends on the ratio of the angular momentum between the inner and outer orbits, with nearly polar orbits preferentially excited for inner sub-Saturn-mass planets over Jupiter-mass planets. This resonance sweeping offers a natural explanation for the excess of sub-Saturns observed to be on polar orbits, regardless of stellar effective temperature \citep{Bourrier2018, Hixenbaugh2023}, as well as a potential mechanism to stunt the runaway growth of sub-Saturns since they are assumed to be embedded within an actively dispersing disk. However, this mechanism predicts similar misalignment rates for compact multi-planet and isolated systems, holding all other factors equal. This would imply that isolated sub-Saturns, which are more often misaligned, host outer companions at a higher rate than sub-Saturns in compact systems, but no statistical studies on these relative rates have yet been completed.

To explain the $T_{\rm eff}-\lambda$ relation and spin-orbit misalignment in single-star exoplanetary systems more broadly, \cite{Hixenbaugh2023} outline a unified model wherein planets gain misalignments if other planets of comparable (or higher) mass form within their disk. Specifically, hotter, more massive stars tend to host more massive protoplanetary disks \citep{Williams2011, Andrews2013, Andrews2020}, which are more likely to form multiple Jupiter-mass planets \citep{Johnson2010, Ghezzi2018, Yang2020}. These Jupiters may interact with each other to produce spin-orbit misalignments after the disk dissipates. By contrast, the limited disk mass around cooler stars is unlikely to produce more than one Jupiter-mass planet, if any \citep{Andrews2013, Ansdell2016, Pascucci2016, Dawson2018, Yang2020}. As a result, these giants, which have no comparable-mass planetary perturbers, should largely retain their primordial alignments throughout the post-disk phase. Nevertheless, despite their less massive disks, cool stars may still form several sub-Saturn planets that undergo an array of dynamical interactions in the post-disk phase that are capable of exciting their stellar obliquities, analogously to multiple Jupiters around hot stars. As present-day compact systems may have avoided violent dynamical interactions, they preserve their primordial spin-orbit angle, allowing them to serve as a window into the early, undisturbed state of planetary systems.

\section*{Acknowledgments}
\par We acknowledge the independent work on the Rossiter-McLaughlin effect measurement of TOI-5398\,b by Mantovan et al. We appreciate the collegiality of the team regarding the manuscript submission process.

We thank the anonymous referee for their careful review of this work and constructive feedback.
We express our gratitude for the insightful discussions with George Zhou and Chelsea Huang regarding the stellar activity of young planetary systems and with Armaan Goyal regarding statistical analyses. 

Data presented were obtained by the NEID spectrograph built by Penn State University and operated at the WIYN Observatory by NOIRLab, under the NN-EXPLORE partnership of the National Aeronautics and Space Administration and the National Science Foundation. These results are based on observations obtained with NEID on the WIYN 3.5m Telescope at Kitt Peak National Observatory (PI: Songhu Wang, IU TAC). Additional NEID data were obtained through time allocated by NSF's NOIRLab (NOIRLab Prop. IDs 2022A-543544, 2023A-904842, 2023B-738235; PI: S. Yee), under the NN-EXPLORE partnership of the National Aeronautics and Space Administration and the National Science Foundation. WIYN is a joint facility of the University of Wisconsin–Madison, Indiana University, NSF's NOIRLab, the Pennsylvania State University, Purdue University, University of California, Irvine, and the University of Missouri. The authors are honored to be permitted to conduct astronomical research on Iolkam Du'ag (Kitt Peak), a mountain with particular significance to the Tohono O'odham.

M.R. acknowledges support from Heising-Simons Foundation grant $\#$2023-4478. S.W. acknowledges support from Heising-Simons Foundation grant $\#$2023- 4050.
We acknowledge support from the NASA Exoplanets Research Program NNH23ZDA001N-XRP (grant $\#$80NSSC24K0153).
This research was supported in part by Lilly Endowment, Inc., through its support for the Indiana University Pervasive Technology Institute.
CP acknowledges support from ANID BASAL project FB210003, FONDECYT Regular grant 1210425, CASSACA grant CCJRF2105, and ANID+REC Convocatoria Nacional subvencion a la instalacion en la Academia convocatoria 2020 PAI77200076.
The Flatiron Institute is a division of the Simons foundation.

All the TESS data used in this paper can be found in MAST: \dataset[10.17909/t9-nmc8-f686]{doi:10.17909/t9-nmc8-f686}.
This research has made use of the NASA Exoplanet Archive \citep{PSCompPars}, which is operated by the California Institute of Technology, under contract with the National Aeronautics and Space Administration under the Exoplanet Exploration Program.

\vspace{5mm}
\facilities{WIYN/NEID}

\software{\texttt{NumPy} \citep{vanDerWalt2011, Harris2020}, \texttt{SciPy} \citep{Virtanen2020}, \texttt{pandas} \citep{McKinney2011pandasAF}, \texttt{matplotlib} \citep{Hunter2007}, \texttt{allesfitter} \citep{Gunther2021}, \texttt{emcee} \citep{Foreman2013}, \texttt{iSpec} \citep{Blanco2014}}, EXOFASTv2 \citep{Eastman2019}



\bibliography{subsaturn}{}
\bibliographystyle{aasjournal}

\end{document}